\def \yskip{\penalty-50\vskip3pt plus 3pt minus 2pt}
\def \reference{\par \yskip \noindent \hangindent .4in \hangafter 1}
\def \abc#1#2#3#4 {\reference#1, {\sl#2}, {\bf#3}, #4}
\def \blank {\lower 5pt\hbox to 0.75in{\hrulefill}}

\def \lae{\mathrel{<\kern-1.0em\lower0.9ex\hbox{$\sim$}}}
\def \gae{\mathrel{>\kern-1.0em\lower0.9ex\hbox{$\sim$}}}

\documentstyle[11pt,aasms,tighten,flushrt]{article}

\begin{document}
\small

\setcounter{page}{1}
\noindent To appear in Asymmetrical Planetary Nebulae II: From Origins to
Microstructures, ASP Conference Series,
J.H. Kastner, N. Soker, \& S. Rappaport, eds.

\begin{center} \bf 
The Transition to Axisymmetrical Mass Loss
\end{center}

\begin{center}
Noam Soker\\
Department of Physics, University of Haifa at Oranim\\
Oranim, Tivon 36006, ISRAEL \\
soker@physics.technion.ac.il 
\end{center}


\begin{center}
\bf Abstract
\end{center}

Any model for the formation of elliptical planetary nebulae (PNs)
should account for the positive correlation between the mass loss
rate and the degree of departure from sphericity of the AGB
progenitor's wind. I propose that this correlation results from
dust formation above magnetic cool spots. The model deals with
elliptical PNs, but not with bipolar PNs. The basic assumption is
that a weak dynamo amplifies magnetic fields in AGB stars, such
that magnetic cool spots are formed, mainly near the equatorial
plane. Enhanced dust formation above these cool spots leads to a
higher mass loss rate in the equatorial plane, resulting in the
formation of an elliptical PN. The dust formation above cool spots
becomes much more efficient when mass loss rate is high. In addition
to explaining the correlation, the model has the advantage that it
can operate for very slowly rotating AGB stars, having rotation
velocities of less than $10^{-4}$ times the break-up velocity. Such
velocities can be achieved by a planet companion of mass
$\sim 0.1 M_{\rm Jupiter}$ which spins-up the envelope, or even
from singly evolved stars which leave the main sequence with a high
rotation velocity. The sporadic nature of magnetic cool spots also
leads to the formation of filaments, arcs, and clumps in the
descendant PN. The model cannot explain the presence of jets and
ansae in elliptical PNs. These are attributed to the destruction
of a planet or brown dwarf on the AGB core. 

\section{Introduction}

 The inner regions of many planetary nebulae (PNs) and proto-PNs
show much larger deviation from sphericity than the outer regions
(for catalogs of PNs and further references see, e.g.,
Acker {\it et al.} 1992; Schwarz, Corradi, \& Melnick 1992;
Manchado {\it et al.} 1996; Sahai \& Trauger 1998;
Hua, Dopita, \& Martinis 1998).
 By ``inner regions'' we refer here to the shell that was formed from
the superwind$-$the intense mass loss episode at the termination of the
AGB, and not to the {\it rim} that was formed by the interaction with the
fast wind blown by the central star during the PN phase
(Frank, Balick, \& Riley 1990).
 This type of structure suggests that there exists a correlation,
albeit not perfect, between the onset of the superwind and the onset
of a more asymmetrical wind.
  In extreme cases the inner region is elliptical while the outer
region (outer shell or halo) is spherical (e.g., NGC 6826, Balick 1987).
 Another indication of this correlation comes from spherical PNs. 
Of the 18 spherical PNs listed by Soker (1997, table 2),
$\sim 75 \%$ do not have superwind but just an extended spherical halo.

 I consider two types of mechanisms that can in principle cause this
correlation.  In the first type ($\S 2$) a primary process or event
causes both the increase in the mass loss rate and its deviation from
spherical geometry.
 A primary mechanism or event may be external or internal to the star.  
 An external event is a late interaction with a stellar or
substellar companion (Soker 1995; 1997), while an internal
mechanism can be the rapid changes in some of the envelope properties
on the upper AGB due to a high mass loss rate and the rapid decrease of
the extended envelope mass.
 Such changes can lead to mode-switch to nonradial oscillations
(Soker \& Harpaz 1992), or an increase in magnetic activity when the 
density profile below the photosphere becomes much shallower,
and the entropy profile much steeper (Soker \& Harpaz 1999).
  In the second type ($\S 3$), the increase in the mass loss rate
on the upper AGB (the so called superwind) makes possible a mechanism which
is very inefficient at low mass loss rates (Soker 2000).

In the present paper I review four recent works on this topic,
all deal with enhanced dust formation, hence enhanced
mass loss rate, above magnetic cool spots
on the surface of AGB stars (Soker 1998; Soker \& Clayton 1999;
Soker\& Harpaz 1999; Soker 2000).
 The mechanisms proposed here do not invoke any new mass loss
mechanism, but use the generally accepted model for the high mass
loss rate on the upper AGB, which includes strong stellar pulsations
coupled with large quantities of dust formation at a few stellar
radii around the stellar surface (e.g., Wood 1979; Jura 1986; Knapp 1986; 
Bedijn 1988;  Bowen \& Willson 1991; Fleischer, Gauger, \& Sedlmayr 1992;
Woitke, Goeres, \& Sedlmayr 1996; Habing 1996; H\"ofner \& Dorfi 1997;
Andersen, Loidl, \& H\"ofner 1999).
 Again, the proposed mechanism(s) applies (apply) only to elliptical PNs,
and not to bipolar PNs. The latter require stellar companions.

\section{Magnetic Cool Spots}

In the first paper (Soker 1998), I presented the basic ingredient of the
model.
The main assumption is that dynamo magnetic activity results in the 
formation of cool spots, above which dust forms much easily.  
The dynamo dictates a general stronger activity toward the equator,
but with significant sporadic behavior.
 The sporadic behavior leads to the formation of filaments, arcs, and
clumps in the descendant PN.
The enhanced magnetic activity toward the equator results in a higher
dust formation rate there, hence higher mass loss rate. 
In that paper I assumed that the dynamo activity increases
as the star ascends the AGB.
Independently, mass loss rate increases as well, due to the
increase of the density scale height (Bedijn 1988; Bowen \& Willson 1991).
 In this model, the increase of dynamo magnetic activity
is attributed to the decreasing density of the envelope, due to
mass loss and expansion, which makes the density profile below the
photosphere much shallower and the entropy profile much steeper
(Soker \& Harpaz 1999).
  The main points of the analysis of Soker (1998) and
Soker \& Harpaz (1999) are as follows.
\newline
(1) In order for the dynamo to stay effective to the upper AGB,
the AGB star should be spun-up by a companion, and/or the dynamo
must be effective even for rotation velocity of
$\omega \sim 10^{-5} \omega_{\rm Kep}$, where $\omega_{\rm Kep}$
is the Kepler velocity of a test particle on the equator. 
 For the envelope spin-up,
if it occurs on the upper AGB, a planet companion of mass
$\sim 0.1 M_{\rm Jupiter}$ is sufficient.
 Born-again AGB stars may hint that the mechanism is efficient even for
$\omega \sim 0.3 \times 10^{-4} \omega_{\rm Kep}$.
\newline
2) The angular velocity decreases rapidly
as the envelope mass decreases toward the termination of the AGB.
In the model this decrease is more than compensated by
the increase of the vulnerability of dust formation and photospheric
conditions to the magnetic activity, due to the shallower density profile
and steeper entropy profile.
\newline
3) The required magnetic activity $\dot E_B$ does not depend on
the mass loss rate.
\newline
4) Because the magnetic energy released through the photosphere
is much below both the kinetic and thermal energy carried by the wind, 
the magnetic activity will not heat the region above the photosphere,
except perhaps in localized regions where the magnetic energy
becomes extremely strong. 
\newline
5) The solar magnetic activity has a cycle of an 11-year period.
If such a cycle exists in upper AGB stars, it will, according to
the proposed model, cause oscillations in the mass loss rate.
 Can it explain the almost periodic shells (or arcs) 
found in several PNs (e.g., CRL 2688 [Egg Nebula], Sahai {\it et al.}
1998; IRAS 17150-3224, Kwok, Su, \& Hrivnak 1998), and the AGB
star IRC+10216 (Mauron \& Huggins 1999)?

\section{Radiation Shielding by Dust}

 In a recent paper (Soker 2000) I consider the second type of mechanism,
where the increase in the departure from spherical mass loss results
from the increase of the mass loss rate. 
 The mass loss rate increases due to the
increase of the density scale height (Bedijn 1988; Bowen \& Willson 1991).
  The large quantities of dust formed above a cool spot during the
high mass loss rate phase shields the region above it
from the stellar radiation.
 This leads to both further dust formation in the shaded region,
and, due to lower temperature and pressure, the convergence of the
stream toward the shaded region, and the formation of a flow
having a higher density than its surroundings.
 This density contrast can be as high as $\sim 4$.
 A concentration of magnetic cool spots toward the equator will lead
to a density contrast of up to $\sim 5$ between the equatorial and polar
directions.
 The shielding does not occur for low mass loss rates, hence the
positive correlation between mass loss rate and the degree of the
departure from sphericity.

 An interesting result of the dust shielding is the required spot size. 
Without shielding, the temperature above a cool spot does not fall 
with radial distance from the surface as steeply as the temperature 
of the environment (Frank 1995). 
 For the region to stay cool enough to form dust, the spot must be large: 
its radius should be $b_s \gtrsim 0.5 R_\ast$ (Frank 1995). 
 This is a large spot, which is not easy to form by concentration of
small magnetic flux tubes (Soker \& Clayton 1999). 
 However, with the dust forming very close to the surface, as is suggested
for cool magnetic spots (Soker \& Clayton 1999), the shielded region forms
dust, which in turn shields a region farther away. 
 Therefore, when dust forms very close to the surface of a small cool spot
with high optical depth, dust will be formed in the entire shaded region 
even when the spot is much smaller than what is required without dust
shielding.
  Not only does the proposed flow allow higher mass loss rate from small
cool spots, it is also limited to small spots.
 Above a large spot the relative mass entering from the surroundings is
small, and since the radiation from the spot is weaker, the material
in the shadow will not be accelerated much.
           
{\bf ACKNOWLEDGMENTS:} 
 This research was supported in part by a grant from
the Israel Science Foundation.

\end{document}